\documentclass[aps,pre,reprint,longbibliography,floatfix,superscriptaddress]{revtex4-2}
\usepackage{graphicx}
\usepackage{amsmath,amssymb,amsfonts}
\usepackage{hyperref}

\begin{document}

\title{Confinement effects on the random sequential adsorption packings of elongated particles in a slit}
\author{Nikolai I. Lebovka}
\email[Corresponding author: ]{lebovka@gmail.com}
\affiliation{Laboratory of Physical Chemistry of Disperse Minerals, F. D. Ovcharenko Institute of Biocolloidal Chemistry, NAS of Ukraine, Kyiv 03142, Ukraine}
%\affiliation{Department of Physics, Taras Shevchenko Kiev National University, Kyiv 01033, Ukraine}

\author{Mykhailo O. Tatochenko}
\email{tatochenkomihail@gmail.com}

\author{Nikolai V. Vygornitskii}
\email{vygornv@gmail.com}
\affiliation{Laboratory of Physical Chemistry of Disperse Minerals, F. D. Ovcharenko Institute of Biocolloidal Chemistry, NAS of Ukraine, Kyiv 03142, Ukraine}

\author{Yuri Yu. Tarasevich}
\email[Corresponding author: ]{tarasevich@asu.edu.ru}
\affiliation{Laboratory of Mathematical Modeling, Astrakhan State University, Astrakhan 414056, Russia}

\date{\today}

\begin{abstract}
The behavior of a system of two-dimensional elongated particles (discorectangles) packed into a slit between two parallel walls was analyzed using a simulation approach. The packings were produced using the random sequential adsorption model with continuous positional and orientational degrees of freedom. The aspect ratio (length-to-width ratio, $\varepsilon=l/d$) of the particles was varied within the range $\varepsilon \in [1;32]$ while the distance between the walls was varied within the range $h/d\in [1;80]$. The properties of the deposits when in the jammed state (the coverage, the order parameter, and the long-range (percolation) connectivity between particles) were studied numerically. The values of $\varepsilon$ and $h$ significantly affected the structure of the packings and the percolation connectivity. In particular, the observed nontrivial dependencies of the jamming coverage $\varphi(\varepsilon)$ or $\varphi(h)$ were explained by the interplay of the different geometrical factors related to confinement, particle orientation degrees of freedom and excluded volume effects.
\end{abstract}

\maketitle

\section{Introduction\label{sec:intro}}

In recent years, the different patterns of self-assembly occuring in random packings of elongated particles onto two-dimensional (2D) substrates have attracted a great deal of attention~\cite{Jahanmahin2020}. Such assemblies may have attractive practical applications in memory devices, the preparation of substrates with controlled wettability, transparent electrodes for optoelectronics, and for sensing materials~\cite{Fang2020,Deng2020}.

Self-assembly can be manifested in a random sequential adsorption (RSA) process consisting of the  irreversible deposition of immobile particles onto a 2D substrate~\cite{Evans1993,Adamczyk2012,Lebovka2020Review}. In the RSA process, the particles are deposited sequentially, their overlapping is forbidden, and above the saturated coverage concentration (``jamming limit'') the deposition process is terminated. For disks, the saturated coverage is $\varphi=0.5472\pm0.0002$~\cite{Hinrichsen1986,Hinrichsen1990}. This value is noticeably smaller than the maximum packing fraction for a close-packed configuration of disks, $\pi/\sqrt{12}\approx0.9069$~\cite{Zong2008}. Moreover, the RSA configuration is not stable  since some disks may be rearranged to create holes sufficiently large to accommodate new disks~\cite{Hinrichsen1990}.

For elongated particles, RSA is a challenging problem that has been the ongoing focus of many researchers. The RSA packings of ellipses~\cite{Talbot2000,Sherwood1990}, rectangles~\cite{Vigil1989,Vigil1990}, discorectangles~\cite{Haiduk2018,Lebovka2020b},  and other elongated  objects~\cite{Perino2017,Budinski-Petkovic2016,Lebovka2020Review,Ciesla2013,Ciesla2013a,Ciesla2015,Lebovka2021} with different aspect ratios (length-to-width ratios $\varepsilon=l/d$), have been analyzed.  In many cases, intriguing non-monotonic $\varphi(\varepsilon)$ dependencies have been observed. Such behavior was explained by a competition between the effects of orientational degrees of freedom and excluded area effects~\cite{Donev2004}.

Spatially confined systems are of special interest for analysis of the effects of particle shape and the effect of confinement on particle self-assembly. The confinement effects have been extensively studied in 2D systems at equilibrium, when enclosed between two parallel hard walls (in a slit), or in square or circular cavities~\cite{Heras2009,DeLasHeras2009,Chen2013,Gonzalez-Pinto2013,Geigenfeind2015,DeLasHeras2014,Sitta2016,Sitta2018,Yao2018}. The behavior of systems of particles with different geometries (ellipses, discorectangles, and rectangles) have also been analyzed~\cite{Heras2013}. In particular, the surface phase diagram of a 2D hard-rectangle fluid confined between two walls was calculated~\cite{Martinez-Raton2007}. Capillary columnar ordering and layering transitions were observed in this system. The density and the order-parameter profiles were calculated and the formation of stationary texture consisting of layers of particles oriented parallel to the walls was demonstrated. Monte Carlo simulation of a system of hard rectangles confined between two walls has been performed~\cite{Triplett2008}. It was shown that confined particles tend to align their long axes parallel to the confining walls with  the effects being more pronounced for smaller separations between those walls.

Strongly confined systems between two parallel walls have been examined, theoretically, for rectangular particles~\cite{Aliabadi2015,Gurin2017,Aliabadi2018}. The density profiles showed planar ordering and damped oscillatory behavior~\cite{Aliabadi2015}. For non-mesogenic particles with small aspect ratios ($\varepsilon < 3$) in the extreme confinement limit~--- small distances between the walls ($h/d\leq 2$), a structural transition from a planar (particle's long axis parallel to the walls) to a homeotropic (particle's long axis perpendicular to the walls) layer with increasing density was observed~\cite{Aliabadi2018}.  For hard discorectangles between the two parallel walls in strongly confined systems ($1<h/d\leq 2$), a rich phase behavior with dependence on the value of the particles' aspect ratio has been observed~\cite{Basurto2020}. For hard ellipses in a circular cavity, the formation of oriented layers in the vicinity of the wall was also reported~\cite{Hashemi2019}.

For the case of confined RSA systems, we are only aware of the study of RSA configurations of hard-discs with diameter $d$ in a narrow slit of width $h$ ($1<h/d<2$)~\cite{Suh2008}. The case $h=d$ corresponds to the one-dimensional (1D) packing of disks onto the line. 1D jamming coverage for lines was  exactly evaluated to be about $\varphi_R=0.7476\dots$ (R\'{e}nyi's parking constant)~\cite{Renyi1963}.

For 1D RSA disc systems, this gives $\varphi=(\pi/4)\varphi_R\approx0.5872$. With an increase of $h$ value, a noticeable decrease of the $\varphi$ value was observed. However, to the best of our knowledge, confined RSA system with elongated particles have never been studied before.

The present work deals with 2D RSA packings of elongated particles (discorectangles) in confined geometry between two parallel plates (a slit). A computationally efficient technique to generate jamming configurations has been employed. The profiles of the coverage and orientational structures of the packings were evidenced by the presence of surface-induced ordering inside the slit. Mean values of the coverage $\varphi(\varepsilon)$ and the order parameter $S$ demonstrated non-monotonic behavior in their dependence on $\varepsilon$ and $h$. The connectivity of particles having a hard core---soft shell structure was also analyzed. The rest of the paper is organized as follows. In Sec.~\ref{sec:methods}, the technical details of the simulations are described and all necessary quantities are defined. Section~\ref{sec:results} presents our principal findings and discussions. Finally, Section~\ref{sec:conclusion} presents some concluding remarks.

\section{Computational model\label{sec:methods}}
\begin{figure}[!htbp]
\centering
\includegraphics[width=0.75\columnwidth]{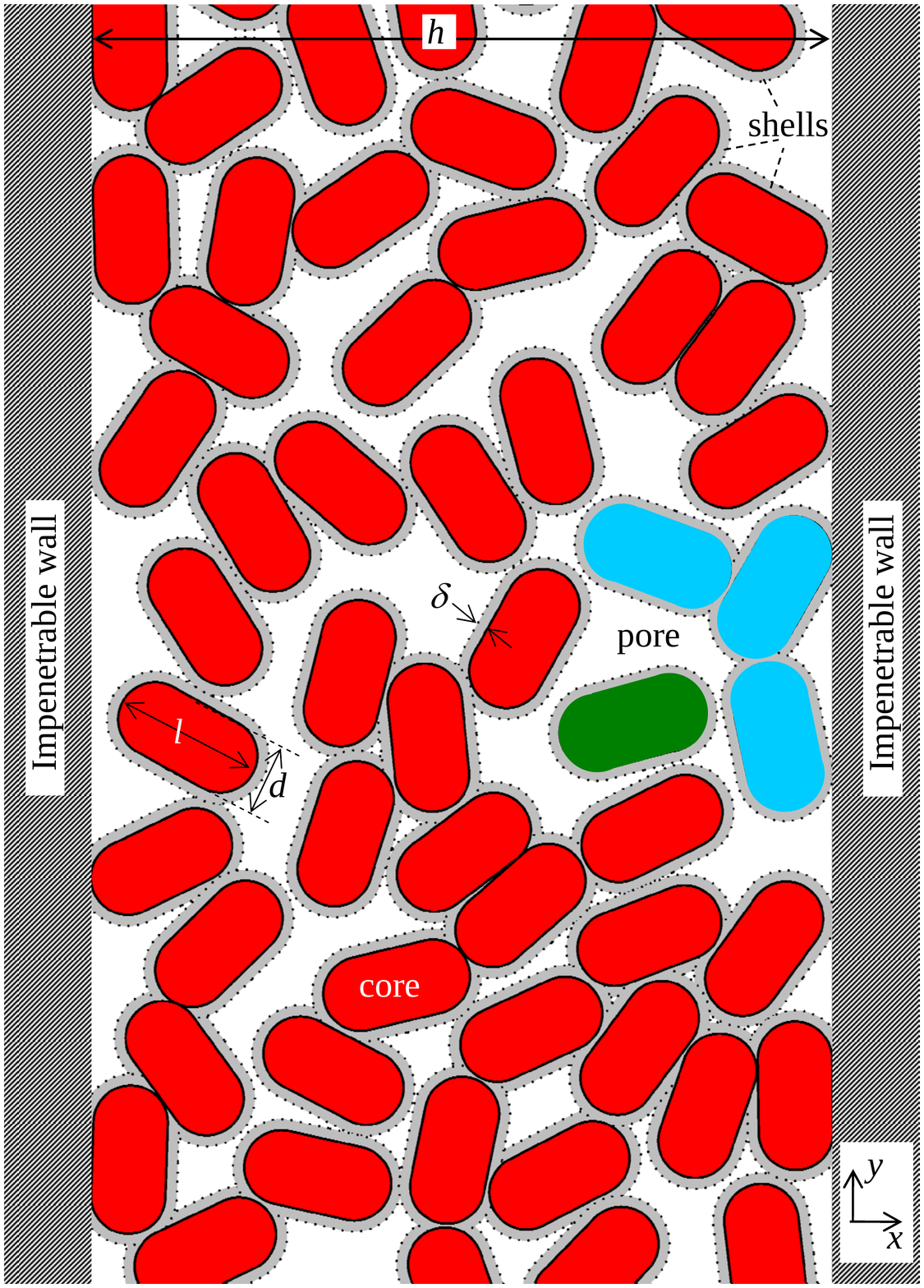}
\caption{Schematic picture of the RSA packing of elongated particles between two parallel walls a  the distance $h$ between them. The walls and particles (cores) are assumed to be impenetrable. The cores of the particles are discorectangles of length $l$ and width $d$. The spaces between particles can be treated as pores. Periodic boundary conditions are applied in $y$ direction. For connectivity analysis along the $x$ and the $y$ axes, the particles were covered with shells of thickness $\delta$. The particles of different shades  correspond to the different clusters. 
\label{fig:f01}}
\end{figure}

Elongated particles were represented by hard discorectangles, which consist of a rectangular part (length $l-d$ and width $d$) with two semicircular caps of diameter $d$ at their opposite ends. To simplify presentation, in further consideration, all lengths are given in units of $d$. Particles with $\varepsilon \in [1;32]$ were analyzed. An RSA model was used for the formation of the packings. Particles with random orientations were randomly and sequentially deposited into a slit (in the space between two parallel impenetrable walls with a distance $h$ between them).  Overlapping of a particle with any previously deposited ones or with the walls was strictly forbidden (Fig.~\ref{fig:f01}). The jamming state was reached when no additional particle could be added to the system due to the absence of any pores of appropriate size. The closest distance of approach of a particle to any of the walls was 0.5.
%%%%%%%%%%%%%%%%%%%%%%%%%%%%%%%%%%%%%%%%%%%%%%%%%%%%%%%%%%%%%%%%%%%%%%%%%%%%%%%%%%%%%%%%%%%%%%%%%%%%%%%%%%%%%%%%%%%%%%%%%%%%%%%%%%%%%%%
%%%%%%%%%%%%%%%%%%%%%%%%%%%%%%%%%%%%%%%%%%%%%%%%%%%%%%%%%%%%%%%%%%%%%%%%%%%%%%%%%%%%%%%%%%%%%%%%%%%%%%%%%%%%%%%%%%%%%%%%%%%%%%%%%%%%%%%

A computationally efficient technique to generate jamming configurations, based on the tracking of local regions, was employed~\cite{Haiduk2018,Lebovka2020b}. The saturated coverage was evaluated as $\varphi=NA_p/A$, where $N$ is the number of deposited particles, and $A_p$ and $A$ are the area of a particle and of the slit, respectively. In the following discussion, only the properties of systems in the jamming state will be analyzed.

The width of the slit was varied within $h \in [1;80]$ and periodic boundary conditions (PBCs) were used in the vertical ($y$) direction.  The comparative data to mimic infinite systems were obtained using PBCs applied in both the horizontal ($x$) and the vertical ($y$) directions using a system size of $32\varepsilon\times32\varepsilon$ (more detailed information can be found elsewhere~\cite{Lebovka2021}).

For connectivity analysis along the $x$ and $y$ axes, a core--shell structure of the particles was assumed. The particle cores were covered with shells of thickness $\delta$ (Fig.~\ref{fig:f01}). The minimum thickness of the overlapping shells, required for the formation of spanning clusters in the $x$ or $y$ direction, was determined using a list of near-neighbor particles~\cite{Marck1997} and the Hoshen---Kopelman algorithm~\cite{Hoshen1976}.

During the deposition of particles with random orientations in a confined slit, some particle orientations may be rejected due to intersection with the walls. Therefore, aligned packing with a preferred orientation along the slit ($y$ direction) is formed. Moreover, for conditions with strong confinement at fairly small value of the slit width $h\leq 2$, ``defective'' packings with diminished density were observed due to a commensuration effect between the width of the particle and the wall separation.

The degree of orientation in the aligned packings that formed was characterized by the order parameter defined as
\begin{equation}\label{eq:S}
  S = \left\langle \cos 2\theta  \right\rangle,
\end{equation}
where $\langle\cdot\rangle$ denotes the average, $\theta$ is the angle between the long axis of the particle and the $y$ axis.

\begin{figure}[!htbp]
	\centering	
\includegraphics[width=0.95\columnwidth]{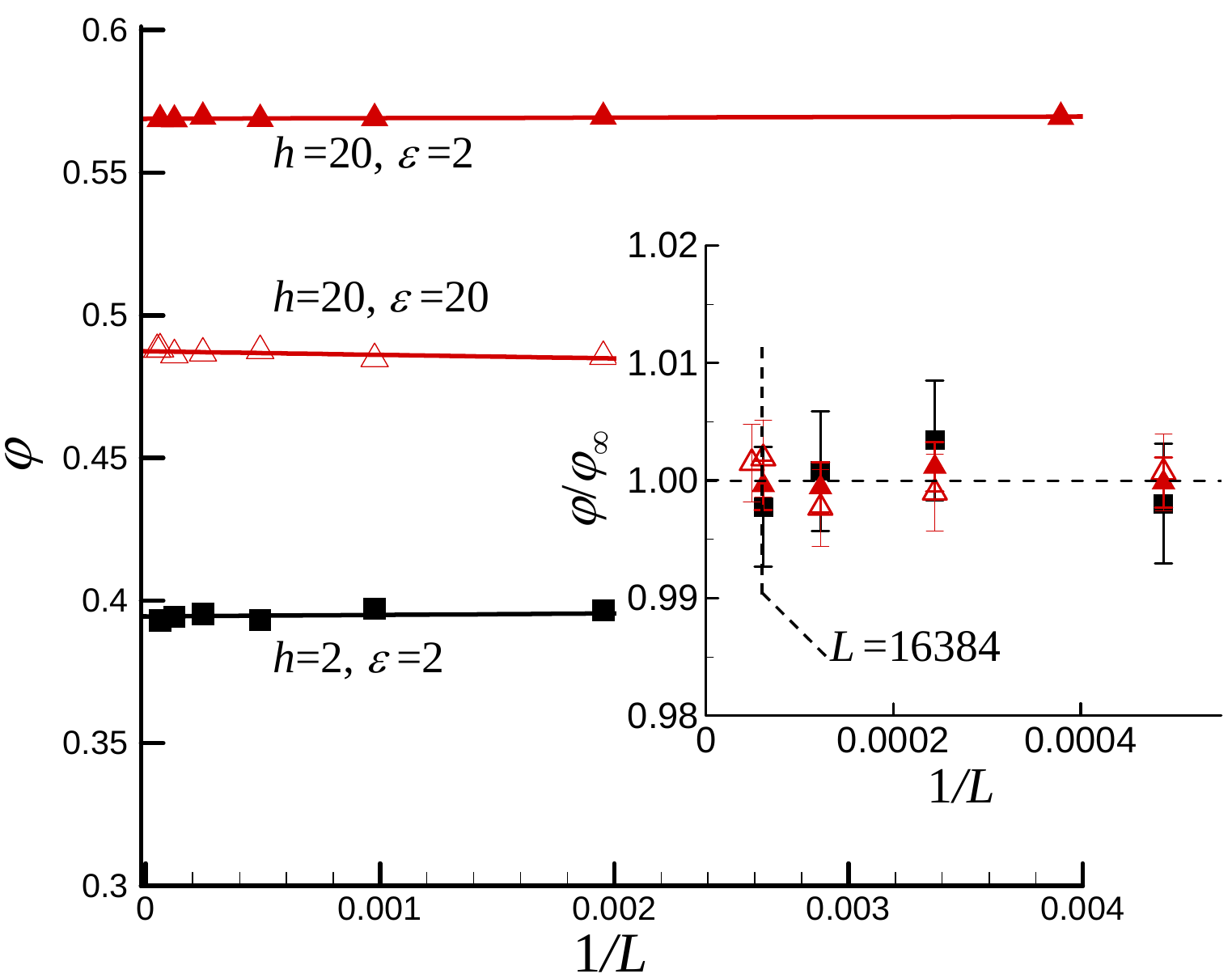}\\
	\caption{Examples of the coverage $\varphi$ versus the inverse length of the slit $1/L$ for  different distances$h$ between its walls and with an aspect ratio of the particles $\varepsilon$. 	Inset shows $\varphi/\varphi_\infty$ versus $1/L$,  where $\varphi_\infty$ corresponds to the limit $L\rightarrow\infty$. \label{fig:f02} }
\end{figure}

The scaling tests with $L= 2^n, n\in [8;14]$ evidenced good convergence of the data for the coverage $\varphi$ (Fig.~\ref{fig:f02}). Therefore, in the present work, the majority of calculations have been performed using $L = 16384, n=14$.
%%%%%%%%%%%%%%%%%%%%%%%%%%%%%%%%%%%%%%%%%%%%%%%%%%%%%%%%%%%%%%%%%%%%%%%%%%%%%%%%%%%%%%%%%%%%%%%%%%%%%%%%%%%%%%%%%%%%%%%%%%%%%%%%%%%%%%%
%%%%%%%%%%%%%%%%%%%%%%%%%%%%%%%%%%%%%%%%%%%%%%%%%%%%%%%%%%%%%%%%%%%%%%%%%%%%%%%%%%%%%%%%%%%%%%%%%%%%%%%%%%%%%%%%%%%%%%%%%%%%%%%%%%%%%%%

Profiles of the local coverage $\varphi (x)$ and the order parameter $S(x)$ were calculated as an ensemble average for the particles that have their centers of mass at distances between $x$ and $x + dx$ from the wall.

For each given value of $\varepsilon$ and $h$, the computer experiments were repeated using 10 to 100 independent runs. The error bars in the figures correspond to the standard deviations of the means. When not shown explicitly, they are of the order of the marker size.

\section{Results and Discussion\label{sec:results}}

\begin{figure*}[!thb]
\centering
\includegraphics[width=0.75\textwidth]{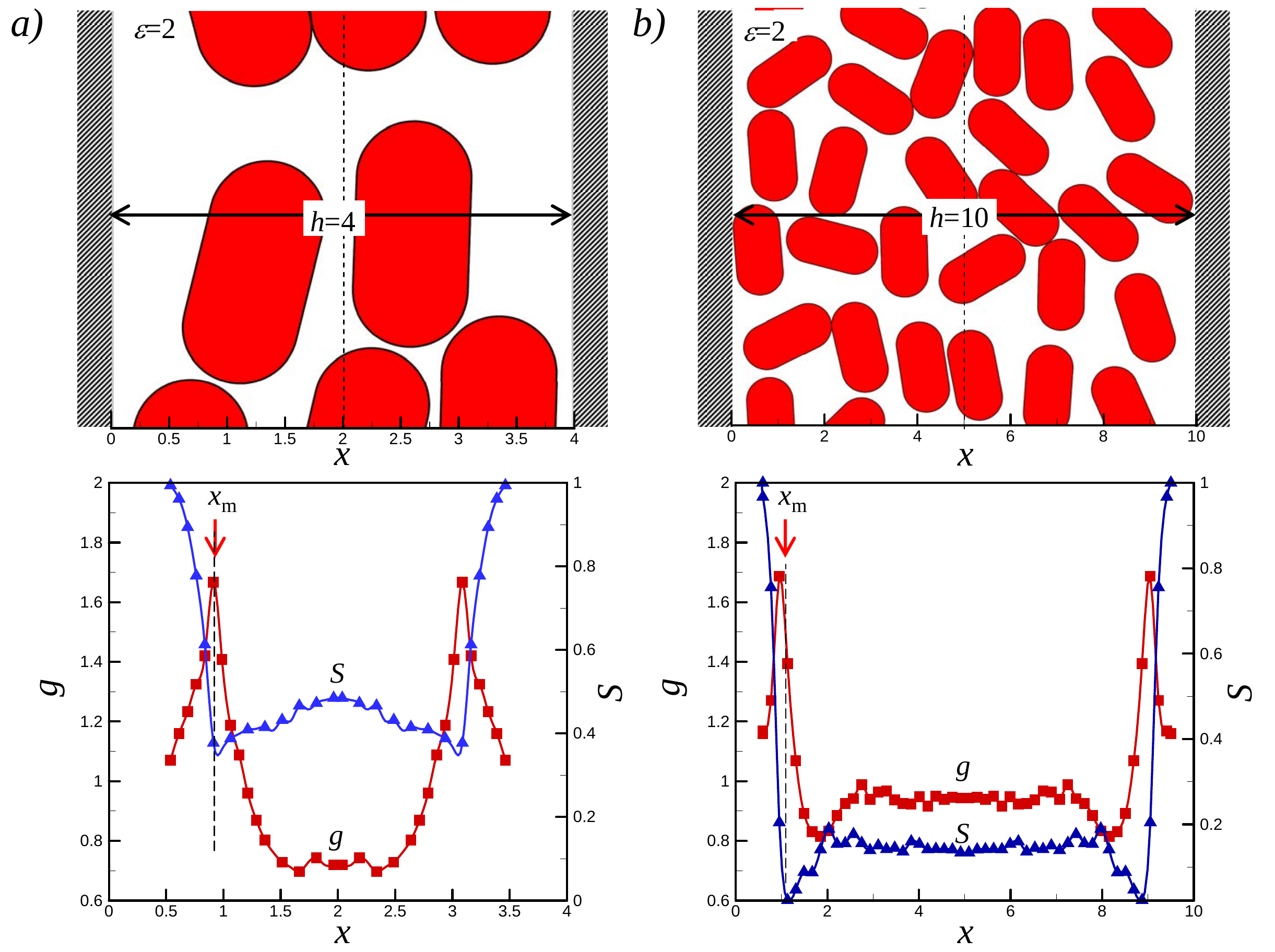}
\caption{Examples of the packing patterns (top), and profiles of the scaled coverage $g(x)=\varphi(x)/\bar{\varphi}$ and the order parameter, $S(x)$ (bottom) for the value of the aspect ratio $\varepsilon=2$ and distances between the walls $h=4$ (a) and $h=10$ (b). The corresponding mean values, averaged across the slit, are $\bar{\varphi}=0.518\pm0.01$ and $\bar{S}=0.54 \pm 0.01$ (a) and $\bar{\varphi}=0.558\pm0.01$ and $\bar{S}=0.26\pm0.01$ (b). Here, $x_\text{m}$  corresponds to the positions of the extremum in the $\varphi^*(x)$ (maxima) or $S(x)$ (minima) dependencies. \label{fig:f03}}
\end{figure*}

Figure~\ref{fig:f03} demonstrates examples of the packing patterns (top), and profiles of the scaled coverage $g(x)=\varphi(x)/\bar{\varphi}$ and the order parameter $S(x)$ (bottom) for the value of the aspect ratio $\varepsilon=2$ and distances between the walls $h=4$ (a) and $h=10$ (b). Here, $\bar{\varphi}$ is the mean value of the coverage, averaged across the slit.
%%%%%%%%%%%%%%%%%%%%%%%%%%%%%%%%%%%%%%%%%%%%%%%%%%%%%%%%%%%%%%%%%%%%%%%%%%%%%%%%%%%%%%%%%%%%%%%%%%%%%%%%%%%%%%%%%%%%%%%%%%%%%%%%%%%%%%%
%%%%%%%%%%%%%%%%%%%%%%%%%%%%%%%%%%%%%%%%%%%%%%%%%%%%%%%%%%%%%%%%%%%%%%%%%%%%%%%%%%%%%%%%%%%%%%%%%%%%%%%%%%%%%%%%%%%%%%%%%%%%%%%%%%%%%%%

The oscillating profiles of $g(x)$ and $S(x)$ were observed for other values of $\varepsilon$ and $h$. Obtained data evidenced that hard walls induced the formation of layers of particles in the vicinity of the walls with the particles' long axes preferentially aligned parallel to the walls.  Both the profiles were symmetrical with respect to a line along the center of the slit.

Similar profiles were captured both by the simulations and from theories for equilibrium systems of particles confined between parallel hard walls with varying separations, $h$. For example, in three-dimensional confined geometries between parallel walls, oscillatory behavior of the local density was observed for hard spheres~\cite{Almarza2009}, spherocylinders~\cite{Mao1997,Dijkstra2001,Heras2006,AghaeiSemiromi2018}, rod-like particles~\cite{Chrzanowska2001}, cylinders~\cite{Malijevsky2010}, and rectangular rods~\cite{Ghazi2020}. Strong structurization of the systems near the walls, and different confined isotropic, nematic, and smectic phases for the elongated particles were observed~\cite{Heras2006}. Similar effects have also been observed in 2D confined geometries between parallel hard walls for ellipses, discorectangles, and rectangles~\cite{Martinez-Raton2007,Triplett2008,Heras2013,Ghazi2020}.

It is remarkable that, in our case for RSA packings using, the selected value of $\varepsilon=2$ (Fig.~\ref{fig:f03}), the coverage and the order parameter profiles revealed a distinctive peak at a distance of about the width of a particle ($x_m\approx1$) from the wall surface. Generally, the positions of the first maximum in the $g(x)$ dependency and the first minimum in $S(x)$ were observed at nearly the same distance from the wall, $x_\text{m}$. The value of $x_\text{m}$ may correspond to the surface layer thickness.

\begin{figure}[!htbp]
	\centering	
\includegraphics[width=0.9\columnwidth]{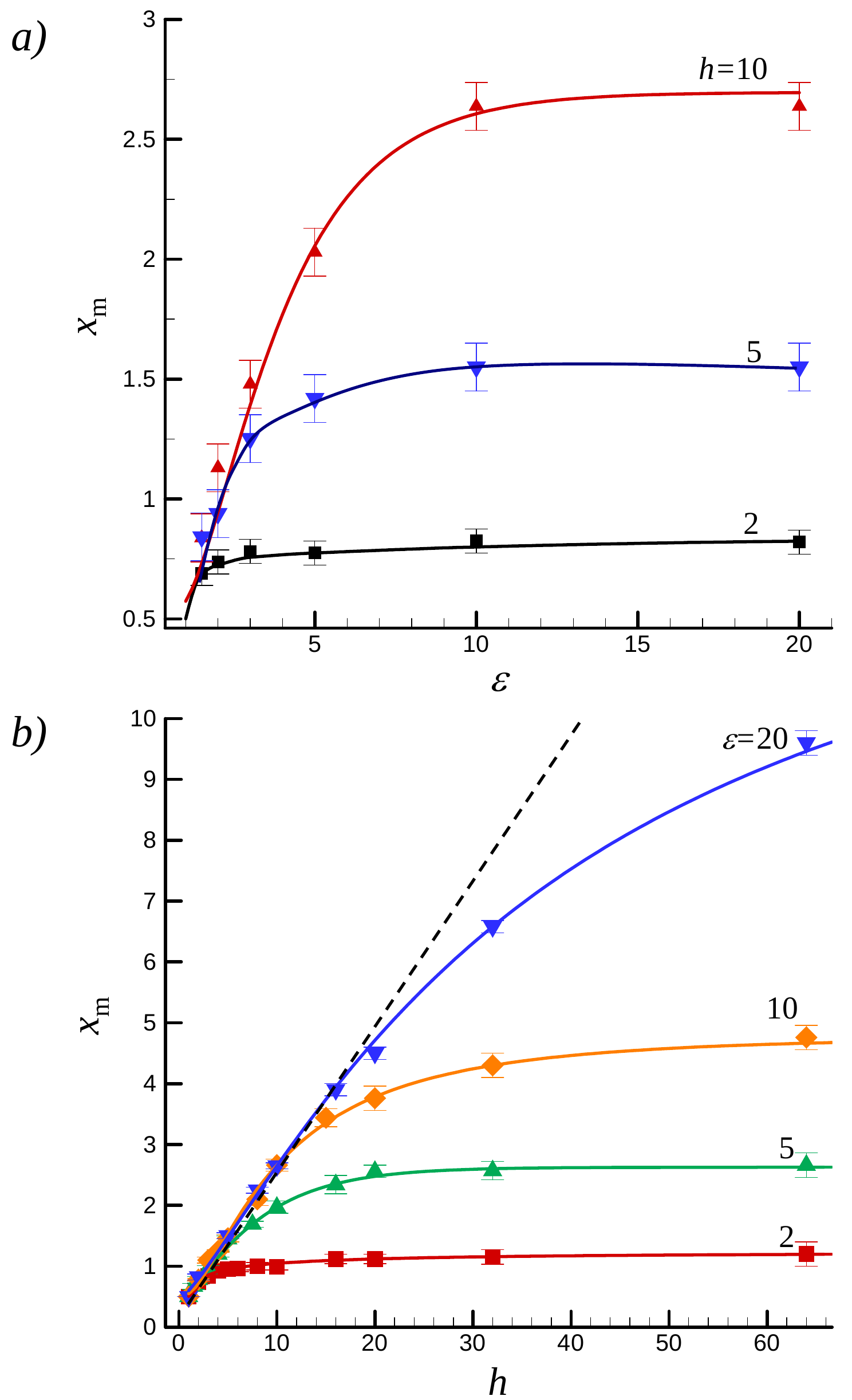}\\
	\caption{Position of the extremum $x_\text{m}$ in the spatial distribution of the coverage $\varphi$ or the order parameter $S$ versus the aspect ratio $\varepsilon$ in a slit with different distances $h$ between the walls (a) and $x_\text{m}$ versus $h$ at different values of $\varepsilon$ (b). \label{fig:f04}}
\end{figure}

Figure~\ref{fig:f04} presents the dependencies of the thickness $x_\text{m}$ versus $\varepsilon$ at different values of $h$ (a) and $x_\text{m}$ versus $h$ at different values of $\varepsilon$ (b). At a fixed value of $h$, the thickness  $x_\text{m}$ initially increased with $\varepsilon$ and saturated for long particles with large values of $\varepsilon$ (Fig.~\ref{fig:f04}a). Specifically, the saturated values of $x_\text{m}$ were $\approx 0.83$ at $h=2$, $\approx 1.54$ at $h=5$, and $\approx 2.56$ at $h=10$. Similarly, at a fixed value of $\varepsilon$, the thickness  $x_\text{m}$ initially increased with $h$ and then saturated for larger values of $h$ (Fig.~\ref{fig:f04}b). The dashed line in Fig.~\ref{fig:f04}b corresponds to the linear dependence $x_\text{m}=0.25(1+h)$. This linear increase of the surface layer thickness with $h$ was only observed in strongly confined systems with $h\leq\varepsilon$, while at larger values of $h$, a  transition of $x_\text{m}(h)$ to saturated behavior was observed. Therefore, in confined systems the extent of perturbation induced by walls depends upon the interrelation between the values of $h$ and $\varepsilon$.
%%%%%%%%%%%%%%%%%%%%%%%%%%%%%%%%%%%%%%%%%%%%%%%%%%%%%%%%%%%%%%%%%%%%%%%%%%%%%%%%%%%%%%%%%%%%%%%%%%%%%%%%%%%%%%%%%%%%%%%%%%%%%%%%%%%%%%%
%%%%%%%%%%%%%%%%%%%%%%%%%%%%%%%%%%%%%%%%%%%%%%%%%%%%%%%%%%%%%%%%%%%%%%%%%%%%%%%%%%%%%%%%%%%%%%%%%%%%%%%%%%%%%%%%%%%%%%%%%%%%%%%%%%%%%%%

\begin{figure}[!htbp]
	\centering	
\includegraphics[width=0.9\columnwidth]{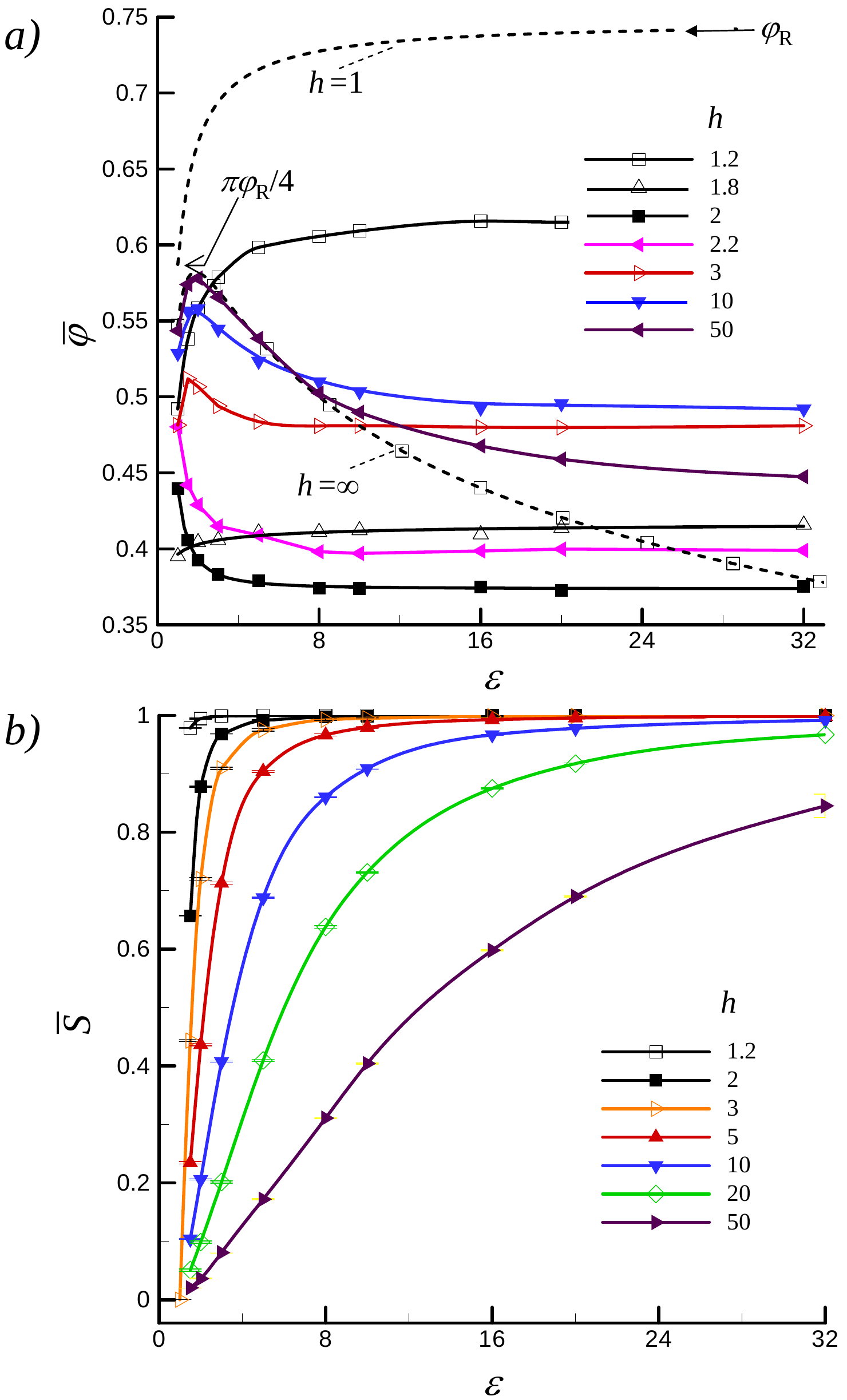}\\
	\caption{Mean coverage $\bar{\varphi}$ (a) and the mean order parameter $\bar{S}$ (b) versus the aspect ratio $\varepsilon$ in slits with different distances $h$ between the walls.
 \label{fig:f05} }
\end{figure}

Figure~\ref{fig:f05} presents the dependencies of the mean coverage $\bar{\varphi}$ (a) and the mean order parameter $\bar{S}$ (b) versus the aspect ratio $\varepsilon$ in slits with different distances $h$ between the walls. For strongly confined 1D systems ($h=1$), the dependence  $\bar{\varphi}(\varepsilon)$ may be calculated as
\begin{equation}\label{eq:icS}
\bar{\varphi}=\varphi_R\frac{\pi/4+\varepsilon-1}{\varepsilon},
\end{equation}
where $\varphi_R=0.7476\dots$ is the R\'{e}nyi's parking constant~\cite{Renyi1963}.
This formula gives the dashed line for $h=1$ presented in Fig.~\ref{fig:f05}a with $\bar{\varphi}=(\pi/4)\varphi_R$ for disks ($\varepsilon=1$), and $\bar{\varphi}=\varphi_R$ for infinitely long particles ($\varepsilon=\infty$).
%%%%%%%%%%%%%%%%%%%%%%%%%%%%%%%%%%%%%%%%%%%%%%%%%%%%%%%%%%%%%%%%%%%%%%%%%%%%%%%%%%%%%%%%%%%%%%%%%%%%%%%%%%%%%%%%%%%%%%%%%%%%%%%%%%%%%%%
%%%%%%%%%%%%%%%%%%%%%%%%%%%%%%%%%%%%%%%%%%%%%%%%%%%%%%%%%%%%%%%%%%%%%%%%%%%%%%%%%%%%%%%%%%%%%%%%%%%%%%%%%%%%%%%%%%%%%%%%%%%%%%%%%%%%%%%

For 2D systems with $h>1$, a wide variety of monotonic and non-monotonic $\bar{\varphi} (\varepsilon)$ dependencies have been observed. Particularly, for the system with PBCs (systems with size $32\varepsilon\times32\varepsilon$ were used to imitate an infinitely large system) the maximum in the $\bar{\varphi} (\varepsilon)$ dependence can be explained a a result the competition between the particles' orientational degrees of freedom and the excluded area effects~\cite{Haiduk2018,Lebovka2020b,Lebovka2021}.

Monotonic $\bar{\varphi} (\varepsilon)$ dependencies have only been observed in strongly confined systems with $h<3$. In particular, confined systems with $h=2$ demonstrated specific behavior with the smallest values of coverage being below $\bar{\varphi}\approx0.4$ for $\varepsilon>2$. However, for higher distances between the walls ($h\geq3$) similar maximums in the $\bar{\varphi} (\varepsilon)$ dependencies have also been detected (Fig.~\ref{fig:f05}a).

The dependencies  $\bar{S} (\varepsilon)$ showed only monotonic character while the order parameter was higher for thinner slits and increased with increase of $\varepsilon$ (Fig.~\ref{fig:f05}b). This behavior reflected the presence of highly oriented layers of particles formed in the vicinity of both walls (Fig.~\ref{fig:f03}). Therefore, confinement by walls resulted in the addition of a  supplementary lever into the competition between the particles' orientational degrees of freedom and the excluded area effects.

\begin{figure}[!htbp]
	\centering	
\includegraphics[width=0.9\columnwidth]{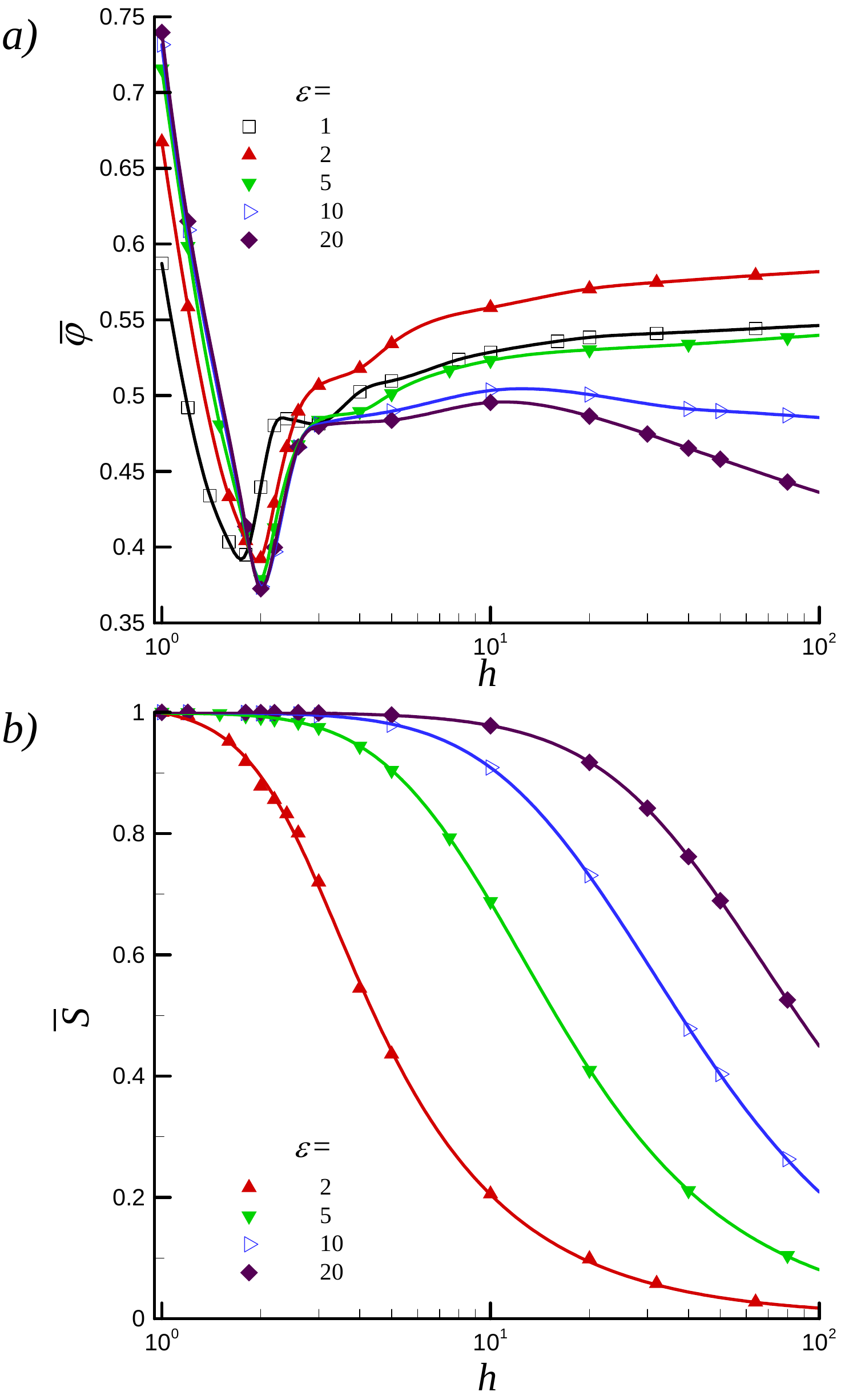}\\
	\caption{Mean coverage $\bar{\varphi}$ (a) and the mean order parameter $\bar{S}$ (b) versus the distance $h$ between the walls for different values of the aspect ratio,~$\varepsilon$.
 \label{fig:f06} }
\end{figure}

Figure~\ref{fig:f06} presents the dependencies of the mean coverage $\bar{\varphi}$ (a) and the mean order parameter $\bar{S}$ (b) versus the slit width $h$ for particles with different values of aspect ratio, $\varepsilon$. The dependencies $\bar{\varphi}(h)$ displayed minimums at approximately the same $h\approx 1.9-2$ separations for all studied values of the aspect ratio, $\varepsilon$ (Fig.~\ref{fig:f06}a). Obtained data for disks  ($\varepsilon=1$) were in good agreement with previously reported data for the range $1<h\leq2$~\cite{Suh2008}.
%%%%%%%%%%%%%%%%%%%%%%%%%%%%%%%%%%%%%%%%%%%%%%%%%%%%%%%%%%%%%%%%%%%%%%%%%%%%%%%%%%%%%%%%%%%%%%%%%%%%%%%%%%%%%%%%%%%%%%%%%%%%%%%%%%%%%%%
%%%%%%%%%%%%%%%%%%%%%%%%%%%%%%%%%%%%%%%%%%%%%%%%%%%%%%%%%%%%%%%%%%%%%%%%%%%%%%%%%%%%%%%%%%%%%%%%%%%%%%%%%%%%%%%%%%%%%%%%%%%%%%%%%%%%%%%

Note, that in strongly confined systems ($1<h\leq 2$) of discorectangles or ellipses, rich phase behavior in the dependence of the value on the particles' aspect ratio has recently been reported~\cite{Basurto2020,Basurto2021}. For small values of the aspect ratio ($\varepsilon=1-5$), the values of the mean coverage $\bar{\varphi}$ showed saturation for larger distances between the walls ($h>10$). However, for particles with large aspect ratio values ($\varepsilon \gtrapprox 10-20$), a decrease in the mean coverage $\bar{\varphi}$ with increase values of $h$ was observed for larger values of the distance between the walls ($h>10-20$). The mean order parameters $\bar{S}$ continuously decreased with increasing values of $h$. The effects were more pronounced for particles with small values of the aspect ratio, $\varepsilon$ (Fig.~\ref{fig:f06}b).

\begin{figure}[!htbp]
	\centering	
\includegraphics[width=0.9\columnwidth]{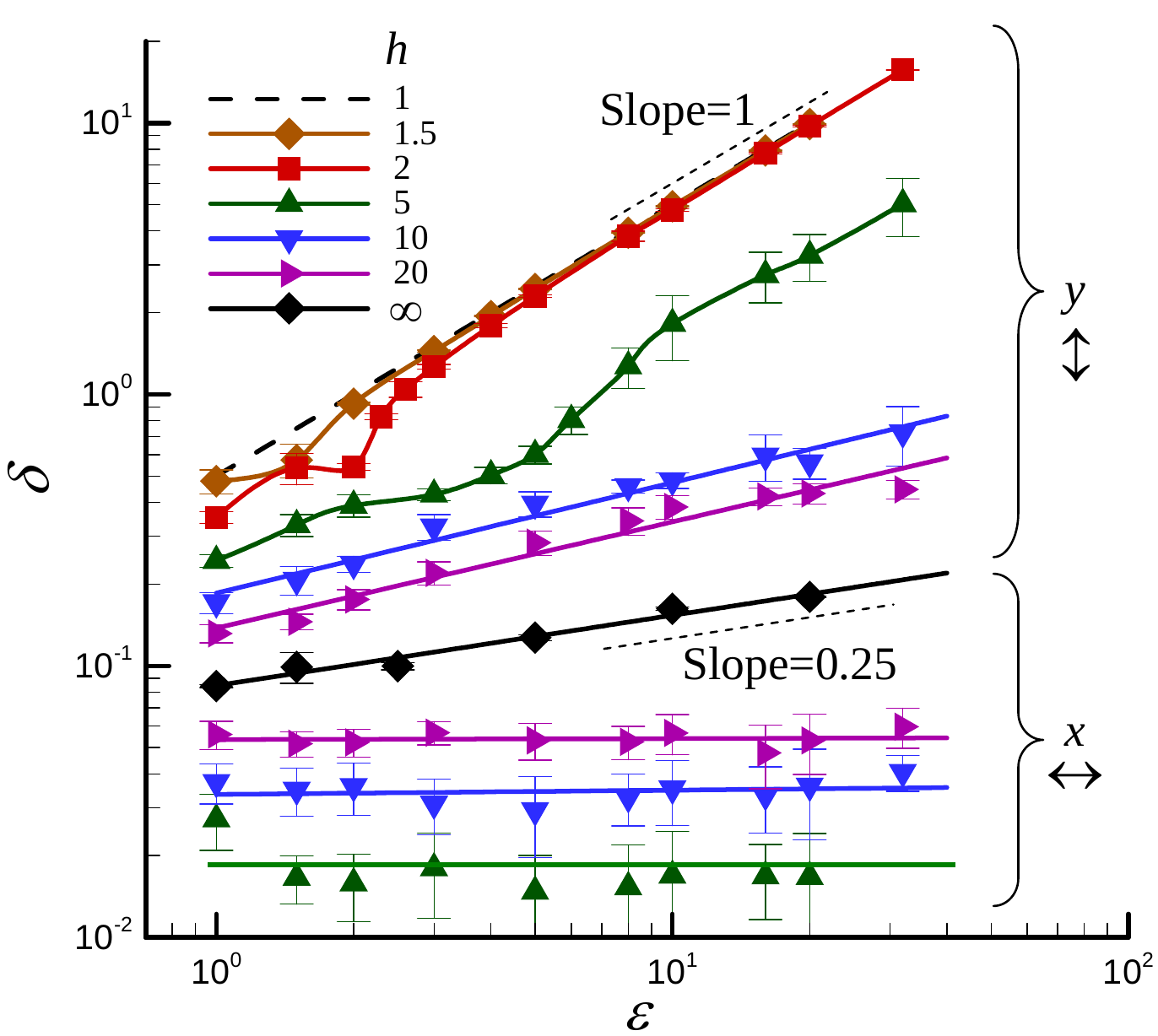}\\
	\caption{Thickness of the connectivity shell $\delta$ versus the aspect ratio $\varepsilon$ along $x (\leftrightarrow)$ and  $y (\updownarrow)$ axes for different distances $h$ between the walls.
 \label{fig:f07} }
\end{figure}

Figure~\ref{fig:f07} presents the thickness of the connectivity shell $\delta$ versus the aspect ratio $\varepsilon$ in directions across the horizontal axis $x (\leftrightarrow)$ and along the vertical axis $y (\updownarrow)$ for different distances $h$ between the walls. The values of
$\delta$  for the horizontal axis $x (\leftrightarrow)$  were relatively small and independent of  the aspect ratio $\varepsilon$.
%%%%%%%%%%%%%%%%%%%%%%%%%%%%%%%%%%%%%%%%%%%%%%%%%%%%%%%%%%%%%%%%%%%%%%%%%%%%%%%%%%%%%%%%%%%%%%%%%%%%%%%%%%%%%%%%%%%%%%%%%%%%%%%%%%%%%%%
%%%%%%%%%%%%%%%%%%%%%%%%%%%%%%%%%%%%%%%%%%%%%%%%%%%%%%%%%%%%%%%%%%%%%%%%%%%%%%%%%%%%%%%%%%%%%%%%%%%%%%%%%%%%%%%%%%%%%%%%%%%%%%%%%%%%%%%

The behavior of $\delta$ along the vertical axis $y (\updownarrow)$ was more complicated. For a 1D-confined system with $h=1$, a linear dependence $\delta=0.5\varepsilon$ along the $y (\updownarrow)$ axis was observed (dashed line with the slope equal to 1). For this case, the connectivity emerged at higher  values for a soft shell thickness around the particles. For strongly confined systems with $h=1.5-5$, the dependencies $\delta (\varepsilon)$ revealed non-monotonic anomalies with inflection points at small values of $\varepsilon$ (Fig.~\ref{fig:f07}). For larger distances ($h\geq10$) between the walls, there were powerlike dependencies $\delta\propto \varepsilon^\alpha$ with exponent $0.25<\alpha<1$. Finally, for the imitated infinite systems (systems with PBCs applied in both the horizontal ($x$) and the vertical ($y$) directions and of the system size $32\varepsilon\times32\varepsilon$~\cite{Lebovka2021}) the value $\alpha\approx0.25$   was estimated. Obtained data evidenced that the confinement between walls resulted in a weakening of connectivity along the vertical axis $y (\updownarrow)$ and an enhancement  of it across the horizontal axis $x (\leftrightarrow)$.

\section{Conclusion\label{sec:conclusion}}

Numerical studies of two-dimensional RSA deposition of discorectangles confined between two parallel walls (in a slit) were carried out. It was shown that the introduction of such confinement significantly affected the distribution of the particles and their orientations inside the slit.
Near the walls, the particles were mostly aligned parallel to the walls. The thickness of the surface layers and the extent of the surface perturbation induced by the walls depend upon the interrelation between the values of $h$ and $\varepsilon$. The mean coverage $\bar{\varphi}$ and the mean order parameter $\bar{S}$ were also strongly dependent upon the values of $h$ and $\varepsilon$.  Particularly, for finite values of $h$, the monotonic $\bar{\varphi} (\varepsilon)$ dependencies have only been observed in strongly confined systems with $h<3$. However, for greater  distances between the walls ($h\geq3$), maxima in the $\bar{\varphi} (\varepsilon)$ dependencies have also been detected. Moreover, minima in the $\bar{\varphi}(h)$ dependencies at approximately the same values of $h\approx 1.9-2$ for all studied values of the aspect ratio, $\varepsilon$, have been observed. This behavior reflects the presence in confined systems of a supplementary lever affecting the competition between the particles' orientational degrees of freedom and the excluded area effects. Obtained data also evidenced that the confinement between walls results in a weakening of the connectivity along the walls and an enhancement of it perpendicular to the walls. Future studies could generalize our approach to different anchoring interactions of particles contained between walls and for confinement within different geometries such as circular or square cavities.

\begin{acknowledgments}
We acknowledge funding from the National research foundation of Ukraine, Grant No.~2020.02/0138 (M.O.T., N.V.V.), the National Academy of Sciences of Ukraine, Project Nos.~7/9/3-f-4-1230-2020,~0120U100226 and~0120U102372/20-N (N.I.L.), and funding from the Foundation for the Advancement of Theoretical Physics and Mathematics ``BASIS'', Grant No. 20-1-1-8-1 (Y.Y.T.).
\end{acknowledgments}

%\cleardoublepage

\bibliography{Lebovka2021_slit}  %

\end{document}